\documentclass[aps,prl,twocolumn,superscriptaddress,groupedaddress]{revtex4}  
\usepackage{graphicx}  
\usepackage{dcolumn}   
\usepackage{bm}       
\usepackage{amssymb}   
\usepackage[section]{placeins}
\usepackage{capt-of}
\usepackage{color}
\usepackage{amsmath}
\usepackage{mhchem}
\usepackage{braket}

\begin{document}

\widetext
\title{Ultrahigh-density spin-polarized H and D observed via magnetization quantum beats}
\author{Dimitris Sofikitis$^{1,2}$, Chrysovalantis S. Kannis$^{1,2}$, Gregoris K. Boulogiannis$^{1,2}$, T. Peter Rakitzis$^{}$}
\email{ptr@iesl.forth.gr}
\affiliation{Institute of Electronic Structure and Lasers, Foundation for Research and Technology-Hellas, 71110 Heraklion-Crete, Greece.}
\affiliation{Department of Physics, University of Crete, 70013 Heraklion-Crete, Greece.}

\date{\today}

\begin{abstract}
	We measure nuclear and electron spin-polarized H and D densities of at least 10$^{19}\, cm^{-3}$ with $\sim$10 ns lifetimes, from the photodissociation of HBr and DI with circularly-polarized UV light pulses. This density is $\sim$6 orders of magnitude higher than that produced by conventional continuous-production methods, and, surprisingly, at least 100 times higher than expected densities for this photodissociation method. We observe the hyperfine quantum beating of the H and D magnetization with a pick-up coil, i.e., the respective 0.7 and 3 ns periodic transfer of polarization from the electrons to the nuclei and back. The $\rm{10^{19}\,cm^{-3}}$ spin-polarized H and D density is sufficient for laser-driven ion acceleration of spin polarized electrons, protons, or deuterons, the preparation of nuclear-spin-polarized molecules, and for the demonstration of spin-polarized D-T or D-$\rm{{^3He}}$ laser fusion, for which a reactivity enhancement of $\rm{\sim50\%}$ is expected.
\end{abstract}

\maketitle

High-density spin-polarized hydrogen (SPH) isotopes are crucial for the measurement of spin-dependent effects in atomic, particle, nuclear, and plasma physics~\cite{PolarizedGas,SpinParticle,PolarizedFusion}. However, many applications are limited or precluded by the inability to produce high densities: polarized laser fusion and laser ion acceleration of hydrogen isotopes require densities of at least $\rm{10^{18}\,cm^{-3}}$ and $\rm{10^{19}\,cm^{-3}}$, respectively. In contrast, conventional methods such as spin-exchange optical pumping (SEOP) or Stern-Gerlach spin separation produce low densities of only $\rm{\sim10^{13}\,cm^{-3}}$~\cite{HighDensityHD,LaserDriven} and $\rm{\sim10^{12}\,cm^{-3}}$~\cite{GasTargets}, respectively. 

Recently, our group has demonstrated the production of highly spin-polarized deuterium (SPD) from the photodissociation of DI with circularly polarized UV light. The ability to produce SPD densities of 10$^{18}$ cm$^{-3}$ with lifetimes of $\sim$1 ns was projected (equivalent to $\sim$10$^{17}$ cm$^{-3}$ with lifetimes of 10 ns), assuming SPH depolarization rates similar to those from collisions with alkali atoms (depolarization from halogen atoms are not known). 

Here, we demonstrate the production of spin-polarized H (SPH) and D (SPD) densities of at least $\rm{10^{19}\,cm^{-3}}$, with $\sim$10 ns lifetimes, from the photodissociation of HBr and DI with  0.15 ns, 213 nm and 266 nm laser pulses, respectively. These SPH densities are at least 100 times higher than projected for depolarization from alkali atoms~\cite{OpticalPumping,LaserPrep,NanoControl,SPD}, as the SPH depolarization rates from halogen atoms are at least 100 times smaller than expected. In addition, we determine that the SPD is depolarized via a DI-D intermediate species, which helps explain the long $\rm{\sim10\,ns}$ polarization lifetimes, as the depolarization rate unexpectedly saturates at high pressures. Therefore, even higher densities are possible. 

The $\rm{10^{19}\,cm^{-3}}$ SPD or SPD densities allow new applications: the demonstration and study of polarized laser fusion~\cite{FusionEngels,SPD} using several $>$5 kJ pulses offered at several laser facilities~\cite{OMEGA,LFEX,Chinese}; polarized ion acceleration with SPH densities of $10^{19}-10^{21}$ cm$^{-3}$ for wavelengths of 10-1 $\mu$m, respectively; maximizing the production of spin-polarized molecules in the NMR detection volume, for signal enhancement requiring gas densities of  $>$$10^{19}$ cm$^{-3}$. 

\begin{figure}[t!]
	\centering
	\includegraphics*[width=0.5\textwidth]{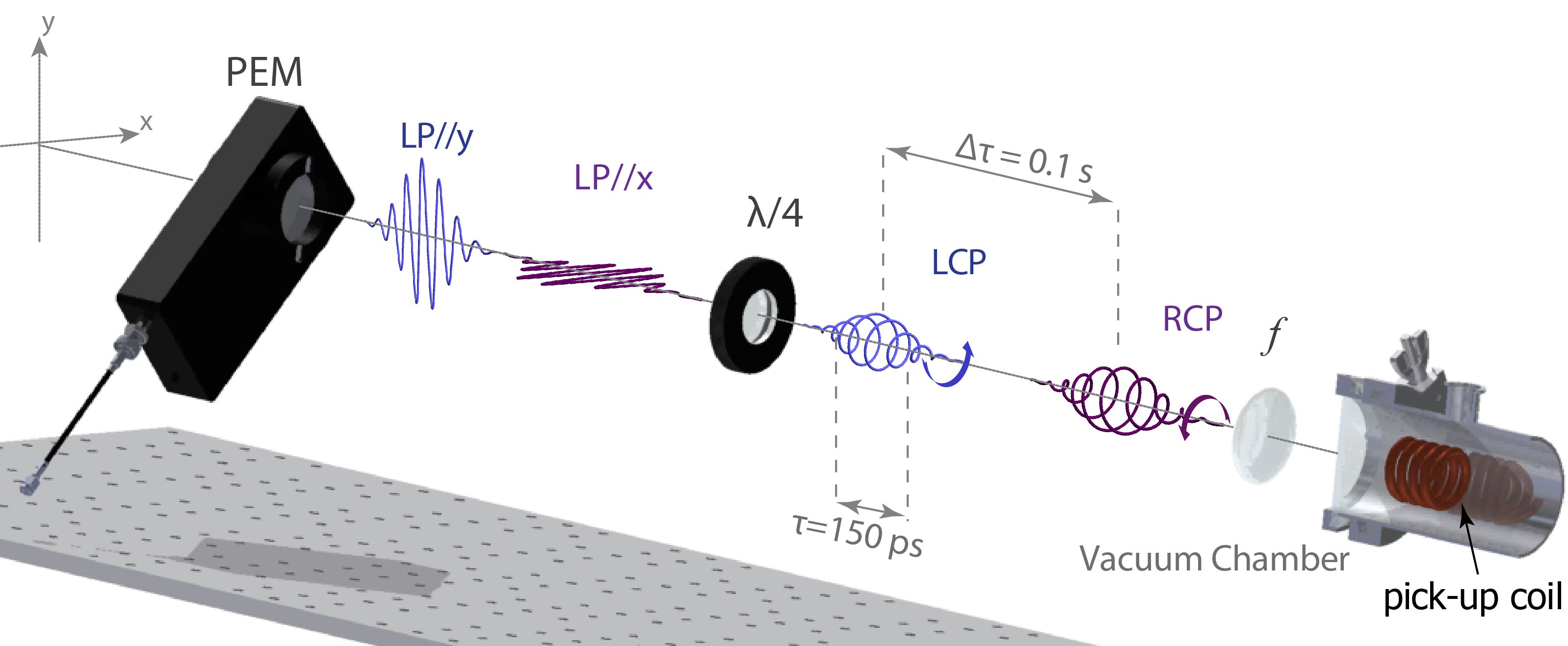}
	\caption{Experimental set-up: Photoelastic modulator (PEM) and quarter-wave plate $\rm{(\lambda/4)}$ alternate laser polarization between Right Circularly Polarized (RCP) and Left Circularly Polarized (LCP) on a shot-to-shot basis at 10 Hz repetition rate. The laser is focused through the pick-up coil, producing a $\rm{\sim2\,mm}$ diameter beam ($f_1\rm{=50\,mm}$), or a focused beam ($f_2\rm{=25\,mm}$, $\rm{4 < r < 200\,\mu m}$).}
	\label{Fig1}
\end{figure}

Our approach to creating ultrahigh densities of spin-polarized atoms, is to take advantage of the ultrafast ($\sim$100 fs) timescales of the UV photodissociation. Such a rapid process generates, nearly instantaneously, highly electronically polarized atoms~\cite{RakitzisScience,RakitzisChemPhys,RakitzisHBr,SofikitisHCl,RydbergDet,ZarePhotodiss,SofikitisOleg} at the original density of the parent molecules, before there is time for depolarization, allowing extremely high densities. Subsequently, the hyperfine interaction transfers polarization from the electronic to the nuclear spin in $\rm{\sim1\,ns}$~\cite{RakitzisPRL}, which is $\rm{\sim6}$ orders of magnitude faster than the polarization step of conventional methods~\cite{HighDensityHD,JetPolarimeter} therefore, highly nuclear-spin-polarized atoms can also be produced at ultrahigh density, on ns timescales. However, spin-polarized atom densities higher than $\rm{10^{12}\,cm^{-3}}$ have not yet been verified from photodissociation, because the optical detection methods used so far fail at high density~\cite{LactateImaging}.

\begin{figure}[htbp!]
	\centering
	\includegraphics*[width=0.5\textwidth]{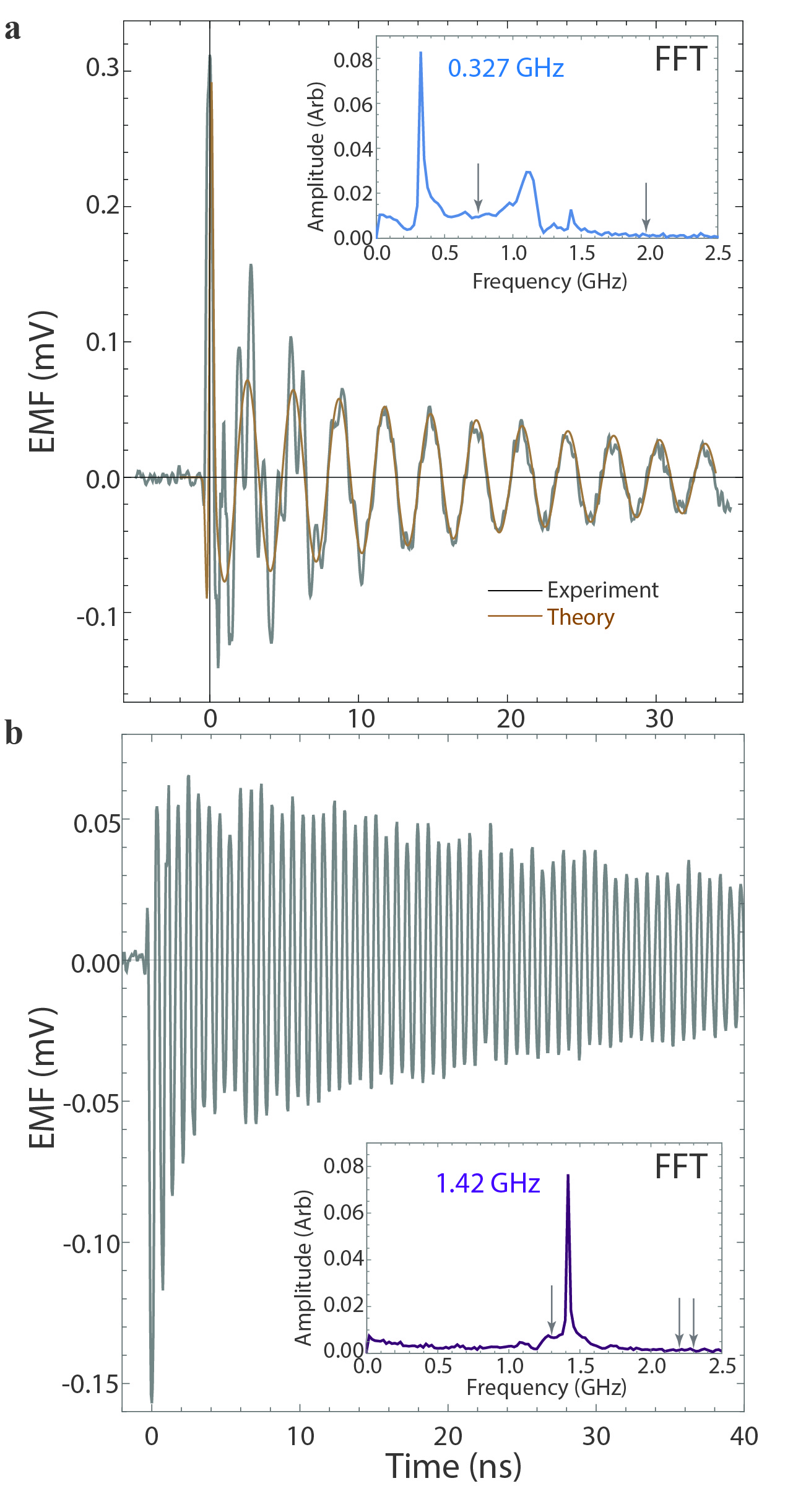}
	\caption{Raw data, theoretical prediction, and FFT of magnetization hyperfine quantum beats. (a) EMF for 125 mbar DI in 875 mbar $\rm{SF_6}$, with $\rm{\tau_p\sim 20\,ns}$. (b) EMF for 70 mbar HBr in 930 mbar $\rm{SF_6}$ with $\rm{\tau_p\sim 40\,ns}$.}
	\label{Fig2}
\end{figure}

The rapid production of large numbers of spin-polarized H or D atoms, and the polarization transfer between the electrons and nuclei, creates a large time-dependent electron magnetization $M_e(t)=2n\braket{m_s(t)}$. The expectation values of the m-state projections for the electronic  $\braket{m_s(t)}$ and nuclear spin $\braket{m_I(t)}$, for H and D, are given, for $t\geq0$, by~\cite{ElectronBeams,RakitzisPRL}:
\begin{equation}
\braket{m_s(t)}=\frac{1}{2}e^{-t/\tau_p}[1-\alpha sin^2(\omega t/2)]
\label{eqExtra}
\end{equation}
\begin{equation}
\braket{m_I(t)}=\frac{1}{2}e^{-t/\tau_p}[\alpha sin^2(\omega t/2)]
\label{eqExtraExtra}
\end{equation}
where $\rm{\alpha=1}$ for H, $\rm{\alpha=32/27}$ for D, and $\omega=2\pi\nu$ is the angular hyperfine frequency for H or D~\cite{RakitzisHBr}. The time-dependent magnetization  creates a current in a pick-up coil (2 mm diameter, 5 mm long, 4.5 turn) through the voltage (EMF), which is used to determine the number of spin-polarized atoms generated by the photodissociation, similar to recent experiments by Milner et al. in the measurement of electron-spin-polarized O$_2$ molecules.~\cite{MilnerPRL}.

We vary the photodissociation laser intensity by several orders of magnitude, by drastically changing the focusing conditions, and show that nearly all of the molecules in the laser focus can be dissociated, leading to densities of $\rm{10^{19}\,cm^{-3}}$. Finally, we study the depolarization-rate dependence on pressure, and elucidate how the dominant depolarization mechanism explains the surprisingly long polarization lifetimes.

The photodissociation of 125 mbar DI at 266 nm produces an EMF trace (Fig.~\ref{Fig2}a) that starts with a sharp peak, from the magnetization produced by the 150 ps rising edge of the laser pulse, followed by a damped oscillation at the deuterium (D) hyperfine frequency with lifetime $\rm{\tau_p \sim 20\,ns}$; the fast Fourier transform (FFT) shows a strong peak at $\rm{\nu_D=327\,MHz}$ (inset). The signal is consistent with $\rm{n=2.5\times10^{13}\,SPD}$ and the expected value of $\rm{p_z(e)=0.24}$~\cite{LaserPrep}. We note that the peak near 1.1 GHz and below in Fig.~\ref{Fig2}a(inset) is due to a large background noise peak related to laser ionization of iodine atoms, which is not completely nulled by the light-polarization subtraction using the PEM. In addition, the peak near 1.4 GHz is caused by SPH from the photodissociation of HI contamination.

A similar damped oscillation, with a higher beating frequency, is observed from photodissociating 70 mbar HBr at 213 nm (Fig.~\ref{Fig2}b); the FFT shows a strong peak at $\rm{\nu_H=1.42\,GHz}$ (inset). There is no evidence in the EMF signals from polarized $\rm{I({{^2P_J})}}$ or $\rm{Br({{^2P_J})}}$ photofragments (arrows in Fig.~\ref{Fig2} insets); we conclude that they are depolarized almost instantly and give no EMF signal.

\begin{figure*}[htbp!]
	\centering
	\includegraphics[width=\textwidth]{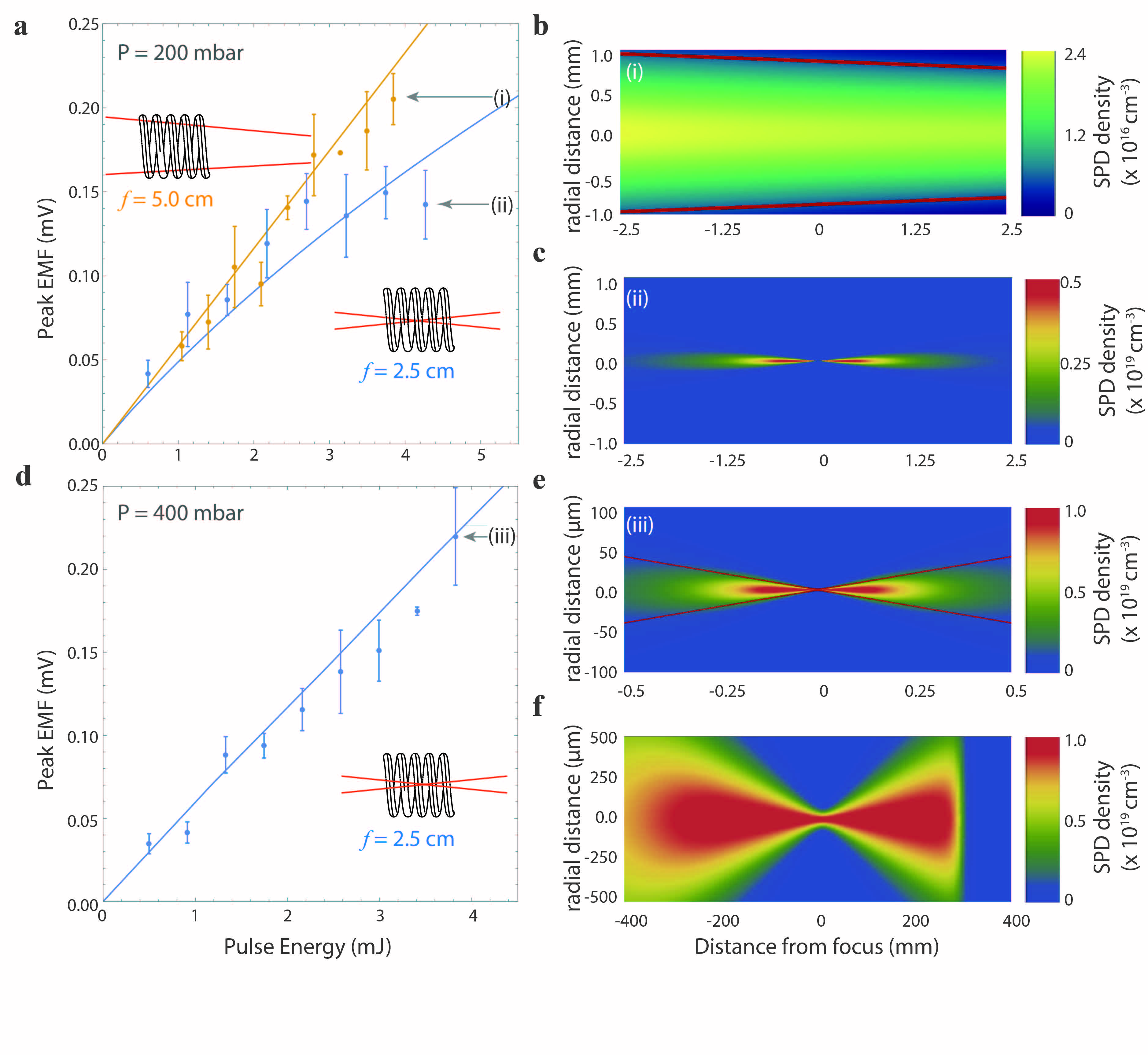}
	\caption{Signal pulse-energy dependence and corresponding SPD density plots. (a) Signal (points) and Beer's-law simulation (lines) of the pulse-energy dependence for 200 mbar DI for: $\rm{\sim2\,mm}$ laser beam (orange); and for focused beam (blue). SPD density plots for: (b) 4 mJ pulse energy and $\rm{\sim2\,mm}$ beam (i), (c) 4.5 mJ pulse energy and focused beam (ii). (d) Signal dependence on pulse energy, for 400 mbar DI and a focused beam (blue circles) and Beer-law simulation (blue line). SPD density plots for (e) 4 mJ pulse energy and a focused beam (iii), and (f) 100 mJ pulse energy and a $\rm{\sim100\,\mu m}$ beam. Error bars are $\rm{2\sigma}$ standard deviation derived from replicate measurements.}
	\label{Fig3}
\end{figure*}

We compare the dependence of the EMF signal on the laser-pulse energy (Fig.~\ref{Fig3}a) for 200 mbar DI, for two focusing conditions. The $\rm{\sim2\,mm}$ laser beam gives an EMF signal that is nearly linear with pulse energy (Fig.~\ref{Fig3}a, orange circles); the focused beam, while yielding signals nearly as large, shows the onset of the saturation of the DI photodissociation (Fig.~\ref{Fig3}b, blue circles). The lines are simulations of the dissociation process using Beer's law, taking saturation into account, and the DI dissociation cross section of $\rm{2\times10^{19}\,cm^2}$ at 266 nm~\cite{OHreaction}. This comparison is made clearer in Fig.~\ref{Fig3}b and~\ref{Fig3}c: the 2 mm beam fills the coil, for a nearly uniform density of $\rm{\sim10^{16}\,SPD\,cm^{-3}}$ (Fig.~\ref{Fig3}b); in contrast, for the focused beam, a similar number of SPD atoms are concentrated in a much smaller volume (Fig.~\ref{Fig3}c), and the SPD density reaches the original parent-molecule density of $\rm{0.5\times10^{19}\,cm^{-3}}$ (red). The good fit of the simulations indicates that ionization or other losses are negligible, within experimental error.
 
Figure~\ref{Fig3}d shows the EMF dependence on the pulse energy (using $f_2\rm{=25\,mm}$) for 400 mbar DI.  The conditions of the experimental point (iii) give the SPD density plot (Fig.~\ref{Fig3}e) which reaches $\rm{10^{19}\,SPD\,cm^{-3}}$ (red). Finally, densities of $\rm{10^{19}\,SPD\,cm^{-3}}$ for volumes of $\rm{\sim0.3\,cm^3}$ can be produced (Fig.~\ref{Fig3}f), in which high-power lasers can be focused for polarized laser-fusion studies.

\begin{figure}[htbp!]
	\centering
	\includegraphics[width=0.5\textwidth]{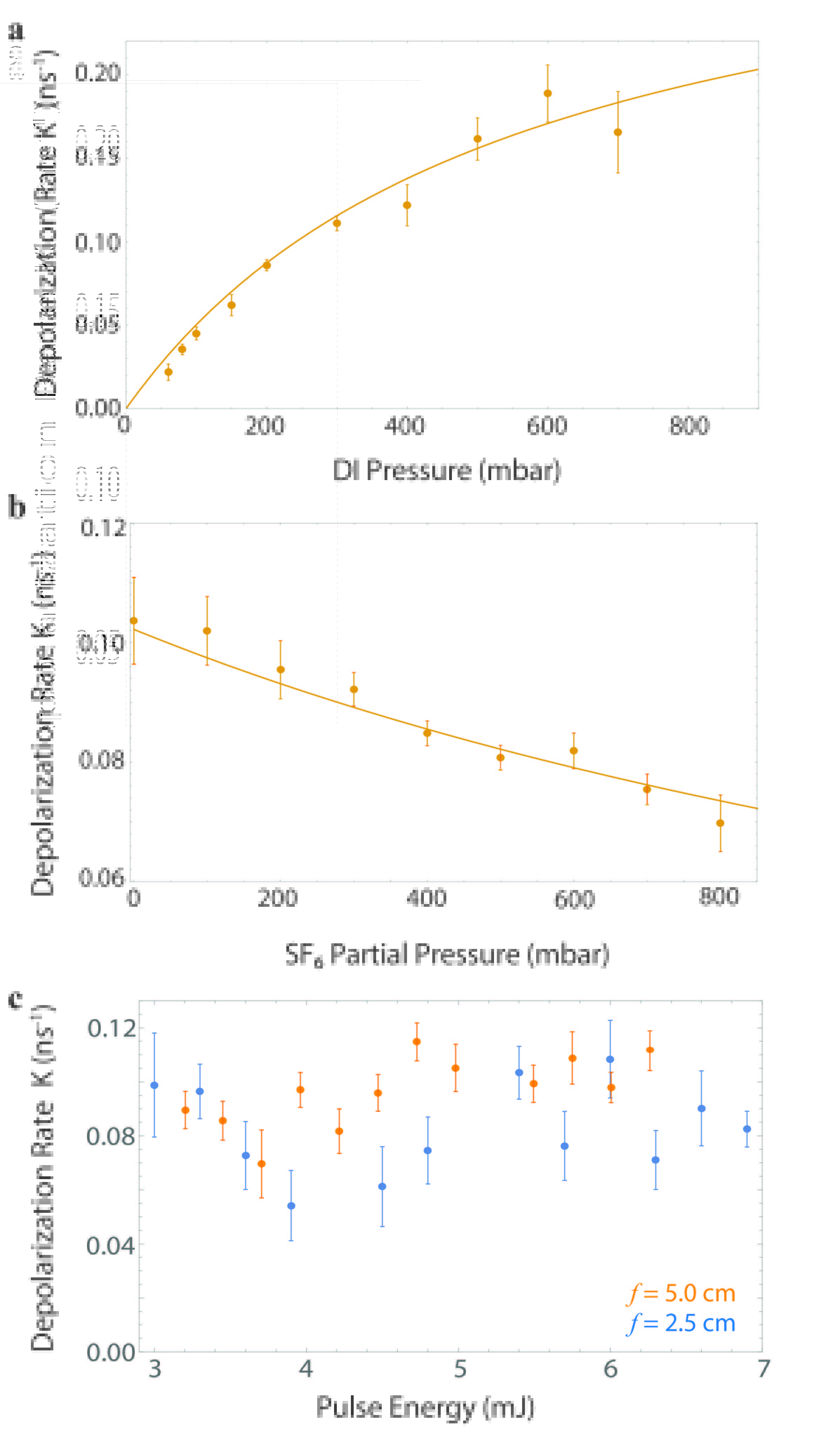}
	\caption{Depolarization-rate dependence. (a) On DI pressure, fit with equation 1. (b) On $\rm{SF_6}$ partial pressure (250 mbar DI), fit with equation 1. (c) On pulse energy (250 mbar DI) for both focusing conditions (Fig.~\ref{Fig1}). Error bars are $\rm{2\sigma}$ standard error from non-linear fit of EMF traces.}
	\label{Fig4}
\end{figure}

We elucidate the SPD depolarization mechanism at high pressure, by measuring the dependence of the $\rm{D^{\uparrow}}$ depolarization rate K=1/$\rm{\tau_p}$ (where $\rm{D^{\uparrow}}$ denotes polarized D atoms) on the DI pressure (Fig.~\ref{Fig4}a). We see a linear dependence at low pressure, which however curves towards a constant value at high pressures. This behavior is not consistent with depolarization caused by DI collisions:
\begin{equation}
\ce{D^{\uparrow} + DI ->[\textit{k}] D + DI}
\label{eq1}
\end{equation}
which predicts a plot of \textit{K} vs. [DI] with slope k, and not the curving behavior in Fig.~\ref{Fig4}a. We find that the simplest process that explains this behavior is the following set of three reactions involving the intermediate molecular species $\rm{DI-D^{\uparrow}}$, formed by collisions of $\rm{D^{\uparrow}}$ with DI:
 \begin{equation}
\ce{D^{\uparrow} + DI <=>[k_1][k_{-1}] DI-D^{\uparrow}}
\label{eq2}
\end{equation}
followed by intramolecular depolarization:
 \begin{equation}
\ce{DI-D^{\uparrow} ->[k_d] DI-D}
\label{eq3}
\end{equation}
or by dissociation of $\rm{DI-D^{\uparrow}}$ via collisions with DI:
 \begin{equation}
\ce{DI-D^{\uparrow} + DI ->[k_2] D^{\uparrow} + 2DI }
\label{eq4}
\end{equation}
or a third body X:
 \begin{equation}
\ce{DI-D^{\uparrow} + X ->[k$^X_2$] D^{\uparrow} + DI + X}
\label{eq5}
\end{equation}
These reactions, assuming the steady-state approximation $(d[\ce{DI-D^{\uparrow}}]/dt=0)$, yield the depolarization rate K:
 \begin{equation}
K=\frac{k_1k_d[DI]}{k_{-1}+k_d+k_2[DI]+k ^X_2[X]}
\label{eq6}
\end{equation}
For $[X]=0$, equation~(\ref{eq6}) fits the data in Fig.~\ref{Fig4}a well, as it predicts linear behavior for small $\rm{[DI]}$, and tends to a constant value ($K=k_1k_d/k_2$) for large $\rm{[DI]}$. Equation~(\ref{eq6}) predicts that the addition of inert gas X, for constant [DI], will decrease the depolarization rate, as [X] is only in the denominator. Indeed, such behavior is shown by $\rm{SF_6}$ in Fig.~\ref{Fig4}b and fit well by equation ~(\ref{eq6}). Therefore, addition of an inert gas lengthens the polarization lifetime, likely allowing even higher densities than $\rm{10^{19}\,cm^{-3}}$. 

Finally, we investigate $\rm{D^{\uparrow}}$ depolarization by the other photodissociation products, $\rm{I(^2P_J)}$ atoms, due to:
\begin{equation}
\ce{D^{\uparrow} + I ->[k^I] D + I^{\uparrow}}
\label{eq7}
\end{equation}
The dependence of \textit{K} on the laser pulse energy is measured, for both focusing conditions, from 3-7 mJ/pulse (Fig.~\ref{Fig4}c). For the highest pulse energies and focusing in the coil, average I-atom densities of $\rm{2\times10^{18}\,cm^{-3}}$ are produced within the laser beam. However, no significant effect on \textit{K} is observed, which allows us to put an upper limit on the I-atom depolarization rate, $k^I<10^{-11}\,cm^3s^{-1}$, which implies that still higher SPD densities are possible than reported here (needing more UV pulse energy).

We propose the production of various nuclear-spin polarized molecules, through SPH reactions, on ns timescales, for pump-probe NMR detection with signal enhancement. In contrast, current methods, such as SEOP~\cite{GasTargets,BouchiatSEOP} and DNP~\cite{DNPlongtime} operate on timescales of minutes or hours, and are less suitable for pump-probe experiments..
The hydrogen isotopes can be easily reacted to form a wide variety of biologically compatible molecules, such as small hydrocarbons and alcohols, as well as radicals. For typical reaction cross section of order ~1 {{\AA}}$^2$, and a (thermalized) speed for the polarized H or D atoms of a few thousand m/s, reactant densities of a few 10$^{19}$ cm$^{-3}$ yield reaction rates of order ~10$^8$ s$^{-1}$, so that all the polarized H and D atoms can react within the polarization lifetime of $\sim$10 ns. Examples of such reactions are given by~\cite{HreactionRates}:

\begin{equation}
\ce{H + C_2H_4 ->[k_3] C_2H_5}\:\mbox{\small$(k_3 = 1.4\times 10^{-12}\frac{cm^3\,s^{-1}}{\,\rm{molecule}})$}
\label{chemicalreaction1}
\end{equation}
\begin{equation}
\ce{H + C_2H_5 ->[k_4] C_2H_6}\:\mbox{\small$(k_4 = 6.0\times 10^{-11}\frac{cm^3\,s^{-1}}{\rm{molecule}})$}
\label{chemicalreaction2}
\end{equation}
with H-atom reaction lifetimes given by 1/k$_3$[C$_2$H$_4$] and 1/k$_4$[C$_2$H$_5$], respectively. The conditions we have demonstrated here are necessary to maximize the NMR signals, as the photodissociation laser pulse will be entirely absorbed within the NMR detection region of length $\approx$1 cm, only for HI and HBr densities in excess of 10$^{19}$ cm$^{-3}$. Furthermore, the SPH atoms can be easily adsorbed at surfaces; this will offer much need sensitivity in NMR of surfaces~\cite{SurfaceNMR}, which can be important for studying complex surface-related processes, such as catalysis~\cite{Catalysis}.

This production of at least $\rm{10^{19}\,SPH/SPD\,cm^{-3}}$ densities is sufficient for the $\lambda$ = 10 $\mu$m laser ion acceleration\cite{OMEGA} of spin-polarized protons, deuterons, or electrons, adding the control of spin to this method of particle acceleration. 

Polarized laser fusion, using densities of at least 10$^{19}$ SPD cm$^{-3}$ and 10$^{20}$ cm$^{-3}$ of polarized $^3$He, and at least 5 kJ/pulse focused to $\sim$10 $\mu$m~\cite{OMEGA,LFEX,Chinese}, will yield well above 10$^4$ neutrons for the D-$^3$He and D-D fusion reactions, needed for polarized fusion studies; well above 10$^6$ neutron will be produced using the 2MJ/pusle NIF laser~\cite{SPD}.

For the D-T or D-$\rm{^3He}$ nuclear fusion reactions, the angular distribution of the neutron or proton products $D(\theta,\phi)$ about the quantization axis, as a function of the nuclear vector polarizations $p_z$ of D, T, or $\rm{^3He}$ is well approximated by~\cite{FusionReactorPlasma}:
\begin{equation}
D(\theta,\phi)=\frac{\sigma_0}{3}[(2+p)-(2p+p_{zz})P_2(cos\theta)]/4\pi
\label{eq10}
\end{equation}
where $p=p_z\rm{(D)}p_z\rm{(Y)}$, $\rm{Y=T}$ or $\rm{^3He}$, $p_z$ is the nuclear vector polarization, $p_{zz}$ is the tensor polarization for D nuclei, $\rm{\sigma_0}$ is the fusion cross section through the intermediate $\rm{^5He}$ or $\rm{^5Li}$ ${\frac{3}{2}^+}$ state for the D-T and D-$\rm{^3He}$ reactions, respectively, and $P_2(x)$ is the $\rm{2^{nd}}$ Legendre Polynomial. The first term in Eq.~(\ref{eq10}) is proportional to the integrated product signal, so that for maximal nuclear polarization, with $p=1$, the product integrated intensity is increased by 50\% compared to $p=0$; also, in this case $p_{zz}=1$, and hence $D(\theta,\phi)\sim 1-P_2(cos\theta)\sim sin^2\theta$.

For the D-$\rm{^3He}$ reaction performed with $p_z(\rm{^3He})=0.76$~\cite{SpinPolarizedMedical,NobleGasNuclei}, $p_z(D,t=1.5\,ns)=0.12$ (reported here), and  $p_{zz}(D)=0$, we predict a 14\% variation in the angular distribution (between $\rm{\theta=0^{\circ}\,and\,90^{\circ}}$) and a 4.5\% increase in the integrated intensity, whereas if bond alignment is used prior to dissociation~\cite{NonResonantLaserFields}, $p_z(D,t=1.5\,ns)=0.5$ can be produced~\cite{LaserPrep}, leading to a 70\% variation in the product angular distribution and a 19\% increase in integrated intensity. In contrast, the effect of polarization in the D-D reaction is poorly understood, with predictions ranging from suppression to enhancement~\cite{PolarizedFusion}. The measurement of the polarized fusion reactions will answer the two outstanding problems of polarized fusion of the past decades~\cite{PolarizedFusion}: determining whether nuclear polarization survives the plasma long enough to benefit fusion reactivity and elucidating the effect of nuclear polarization in the D-D reactions~\cite{NuclearFusionProceedings}.

\textbf{Acknowledgments} This work is supported by the project ``HELLAS-CH" (MIS 5002735) which is implemented under the ``Action for Strengthening Research and Innovation Infrastructures", funded by the Operational Programme ``Competitiveness, Entrepreneurship and Innovation" (NSRF 2014-2020) and co-financed by Greece and the European Union (European Regional Development Fund). We thank S. Pissadakis for access to the EKSPLA SL312M laser, and to D. Vlassopoulos for loan of the PEM. We thank Ralf W. Engels and Andrew Sandorfi for carefully reading the manuscript.

\end{document}